# Accretion/jet activity and narrow [O III] kinematics in young radio galaxies

Qingwen Wu[1,†], Minfeng Gu[2] & Andrew Humphrey[3,1]

1. Korea Astronomy and Space Science Institute, Daejeon 305348, Republic of Korea
2. Key Laboratory for Research in Galaxies and Cosmology, Shanghai Astronomical Observatory, Chinese Academy of Sciences, 80 Nandan Road, Shanghai 200030, China
3. Instituto Nacional de Astrofisica, Optica y Electronica (INAOE), Aptdo. Postal 51 y 216, 72000 Puebla, Pue., Mexico

We estimate black hole masses and Eddington ratios for a sample of 81 young radio galaxies, which includes 42 compact steep-spectrum (CSS) and 39 gigahertz-peaked spectrum (GPS) sources. We find that the average black hole (BH) mass of these young radio galaxies is $<\log M_{bh}> \sim 8.3$, which is less than that of radio loud QSOs and low redshift radio galaxies ($<\log M_{bh}> \sim 9.0$). The CSS/GPS sources have relatively high Eddington ratios, with an average value of $<\log L_{bol}/L_{Edd}> = -0.75$, which are similar to those of narrow line Seyfert 1 galaxies (NLS1s). This suggests that young radio galaxies may not only be in the early stages of their radio activity, but also in the early stage of their accretion activity. We find that the young radio galaxies as a class deviate systematically from $M_{bh}$-$\sigma$ relation defined by nearby inactive galaxies, when using $\sigma_{[O\,III]}$ as a surrogate for stellar velocity dispersion $\sigma$. We also find that the deviation of the [O III] line width, $\Delta\sigma = \sigma_{[O\,III]} - \sigma_{[pred]}$, is correlated with the Eddington ratio; sources with $L_{bol}/L_{Edd} \sim 1$ have the largest deviations, which are similar to those of radio quiet QSOs/NLS1s (i.e., sources in which the radio jets is absent or weak), where $\sigma_{[pred]}$ are calculated from the Tremaine et al. relation using our estimated BH masses. A similar result has been obtained for 9 linear radio Seyfert galaxies. On the basis of these results, we suggest that, in addition to the possible jet-gas interactions, accretion activities may also play an important role in shaping the kinematics of the narrow [O III] line in young radio galaxies.

accretion, accretion discs - ISM: jets and outflows - radio continuum: galaxies - quasars: emission lines

Gigahertz-peaked spectrum (GPS: projected linear size $D<1$ kpc) and compact steep-spectrum (CSS: projected linear size $D<20$ kpc) radio sources constitute a large fraction (~40%) of the powerful radio source population. Their radio spectra are simple and convex with peaks close to 1 GHz and 100 MHz for GPS and CSS sources, respectively[see 1 for a review]. Two basic models have been proposed to explain the CSS/GPS phenomena: (1) *the youth scenario*: the sources are small because they are young, and they will grow to large sized during the course of their life[e.g., 2]; (2) *the frustration scenario*: the radio emitting plasma is permanently confined within the host galaxy by an external dense medium[e.g.,3]. The youth scenario is strongly support by measurements of kinematic[e.g., 4] and radiative age[e.g.,5], which provide values between $\sim 10^2$ and $\sim 10^5$ years. Czerny et al. (2009)[6] proposed that the young radio galaxies may be due to intermittent activity of the active nucleus caused by radiation pressure instability in the accretion disk.

If the evolutionary hypothesis of the youth scenario is correct, study of the early-phase of radio activity and feedback in CSS/GPS sources may impact our understanding of galaxy formation and evolution. Narrow line Seyfert 1 galaxies (NLS1) are a particular class of active galactic nuclei (AGN). The most widely accepted paradigm for NLS1s is that they accrete at close to the Eddington rate and have smaller black hole (BH) masses compared to broad line AGN, and they might be in the early stage of AGN evolution[e.g.,7]. Most of the radio loud NLS1 galaxies have compact, steep-spectrum sources, which are similar to CSS/GPS sources[e.g.,8,9]. Therefore, compact radio sources (CSS/GPS) may not only be in the early stage of radio activity, but also in the early stage of BH accretion.

Received;   accepted
doi:
†Corresponding author (Email:qwwu@shao.ac.cn)
This work is supported by National Science Foundation of China (grants 10633010, 10703009, 10833002 and 10821302), and 973 Program (No. 2009CB824800).

There is increasing evidence that narrow line regions of radio loud AGN, especially for young radio galaxies, have very complex dynamical structures. The kpc-scale narrow line regions typically have a similar spatial extent to radio jets in these galaxies (especially in CSS sources). Therefore, shocks caused by the interaction of jet and interstellar medium (ISM) are assumed to play an important role in shaping the dynamics of narrow emission lines in these young radio galaxies[e.g.,10], and some luminous linear radio sources[e.g.,11,12]. The broader line width of [O III] in these young radio galaxies is usually thought to be driven by the expanding radio jets[e.g.,13,14,15,16] (jet-driven outflows). The radio quiet NLS1s also show broader [O III] line width compared to their virial velocity[e.g.,17,18]. However, the width of low-ionization lines (e.g., [S II]) or the core component of [O III] (after removal of asymmetric blue wing) of NLS1s is still a good proxy for stellar velocity dispersion[e.g., 19,20]. Therefore, strong radiation pressure caused by accretion activity (e.g., accretion-driven outflows/winds) may also affect the kinematics of the [O III] emission line[e.g.,19,20], since that the jet is absent or weak in these radio quiet NLS1s. Ionization diagnostics using various narrow lines suggest a mix of shock and photo-ionization to explain the excitation of the extended emission line gas[e.g., 21,15].

Young radio sources contain two potential mechanisms to drive outflows: radio jet and accretion activity. In this work, we present the BH masses and Eddington ratios for a sample of CSS/GPS sources, and then explore the possible relations between the [O III] kinematics and the accretion/jet properties. Throughout this paper, we assume the following cosmology: $H_0$=70 km s$^{-1}$ Mpc$^{-1}$, $\Omega_0$=0.3 and $\Omega_\Lambda$=0.7.

# 1. Sample

The young radio galaxies used in these work are mainly selected from Wu (2009a, and references therein)[22] and Wu (2009b, and references therein)[23], where the BH masses and bolometric luminosities are estimated by various methods. The [O III] kinematics (e.g., full width at half maximum, FWHM) for most of sources is also included in Wu (2009a)[22]. In total, we present 42 CSS sources, of which 38 sources are selected from Wu (2009a)[22], 3 sources are selected from Wu (2009b)[23] and 1 source is selected from Gu et al. (2009)[24]. We also selected 39 GPS sources, of which 27 sources are selected from Wu (2009a)[22] and 12 sources are selected from Wu (2009b)[23]. We note that our sample is a characteristic, rather than complete sample of compact radio sources. In this work, we also include 9 linear radio sources selected from the literature to investigate the possible [O III] kinematics, of which the narrow lines, traditionally, are also believed to be strongly affected by the radio jets[e.g.,11,12]. Table 1 lists the sample of linear radio Seyfert galaxies with the relevant information.

Table 1 The data of linear radio sources.

| Source | z | $M_{bh}$ log($M_{sun}$) | $L_{bol}$ log(erg/s) | $L_{bol}/L_{Edd}$ | $\sigma_{[O III]}$ log(km/s) | Refs.[a] |
|---|---|---|---|---|---|---|
| Mrk 348 | 0.015 | 7.21 | 44.27 | -1.10 | 2.17 | 25,26 |
| Mrk 6 | 0.019 | 7.80 | 44.41 | -1.50 | 2.29 | *[b],26 |
| NGC 1068 | 0.004 | 7.23 | 44.98 | -0.39 | 2.64 | 25,26 |
| Mrk 3 | 0.014 | 8.65 | 44.54 | -2.25 | 2.54 | 25,26 |
| Mrk 78 | 0.037 | 7.87 | 44.59 | -1.42 | 2.64 | 25,27 |
| Mrk 622 | 0.023 | 6.92 | 44.52 | -0.54 | 2.67 | 25,26 |
| NGC 2992 | 0.008 | 7.72 | 43.92 | -1.94 | 2.0 | 25,26 |
| Mrk 1066 | 0.012 | 7.01 | 44.05 | -0.60 | 2.22 | 25,26 |

a) References for $M_{bh}$, $L_{bol}$ (row 1), and $\sigma_{[O III]}$ (row 2) respectively.
b) The $M_{bh}$ and $L_{bol}$ are calculated in this work with broad line width from Wang et al. (2006)[28].

# 2. Results and Discussions

## 2.1 BH mass and Eddington ratio

The masses of CSS/GPS sources are mainly derived from their bulge stellar velocity dispersions $\sigma$, host galaxy luminosities, or the empirical relation between the broad line width and line luminosity[see 22,23]. The BH masses of several CSS sources derived from broad line width and optical continuum are also used: since steep-spectrum radio sources tend to be oriented with jets pointing away from our line of sight, the jet contribution at optical wavelengths is expected to be negligible, due to the weakness of beaming effects.

We find that the BH masses vary by several orders of magnitude, from $10^7$-$10^{10}$ $M_{sun}$, for both CSS and GPS sources. The histogram distributions of the BH masses are shown in Figure 1. We find that the average logarithmic BH masses are <log $M_{bh}$>= 8.34, 8.28, and 8.31, with standard deviations are 0.81, 0.56, and 0.59, for 42 CSS sources, the 39 GPS sources, and the whole sample, respectively. We note that the average BH mass of young radio galaxies (<log $M_{bh}$>~8.3) is systematically less than that of radio loud QSOs (<log $M_{bh}$>~9.1[e.g.,29]) or low redshift radio galaxies (<log $M_{bh}$>~8.9[e.g.,30]).

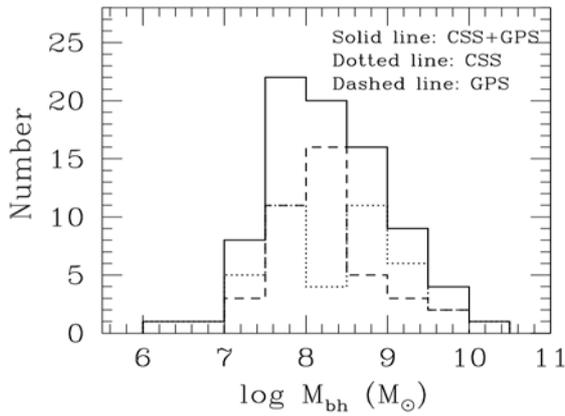

**Figure 1** The histogram distribution of the BH mass.

One of the possible physical reasons is that the BHs in these young radio galaxies are still growing rapidly.

We use the broad emission line luminosity to derive the bolometric luminosity, which is expected to be free from contamination by the nonthermal emission of the jet. The empirical correlations between broad line luminosities and bolometric luminosities for a sample of radio quiet AGN (broad lines $H\beta$, Mg II and C IV) are used to derive the bolometric luminosities of the young radio galaxies in our sample[see 22]. The bolometric luminosities of 12 GPS sources are also estimated from [O III] emission lines, where the [O III] line emission may still be ionized by the accretion processes[see 23]. The reason is that the narrow line regions of these GPS sources (around several kpc in size) have larger linear sizes than the radio jets (<1 kpc), and therefore shocks caused by jet-ISM interaction are unlikely to be important[see also 31]. We find that the bolometric luminosity of GPS source OQ 208 calculated from [O III] line agrees very well with that derived from the broad $H\beta$, and this also supports the idea that the contribution from jet-induced shocks is not important in the GPS sources.

We estimate the Eddington ratios, $L_{bol}/L_{Edd}$, from the estimated BH masses and bolometric luminosities. We find that most of the CSS/GPS sources are high Eddington ratio systems ($10^{-2}<L_{bol}/L_{Edd}$<several, see Figure 2). The average Eddington ratios are <log $L_{bol}/L_{Edd}$>=−0.55, −0.99, and −0.75, with standard deviations are 0.41, 0.98, and 0.2, for the CSS sources, the GPS sources and the whole sample, respectively; these averages are higher than that of broad line radio galaxies <log $L_{bol}/L_{Edd}$>=−1.91 or of radio loud QSOs (<log $L_{bol}/L_{Edd}$ >=−1.11[e.g.,29]), but similar to that of NLS1s (<log $L_{bol}/L_{Edd}$ >~−0.8[e.g.,32]). Our results suggest that these young radio galaxies are also in an early stage of accretion activity, and that the BHs are still growing rapidly. We note that the average Eddington ratio of GPS sources in this work is slightly lower than that of Wu (2009a)[22], most likely due to several GPS radio galaxies being included in this work while only GPS QSOs were included in Wu (2009a)[22]. More uniform and complete sample is still preferred to further constrain the BH masses and Eddington ratios in these young radio sources, which will be our future work.

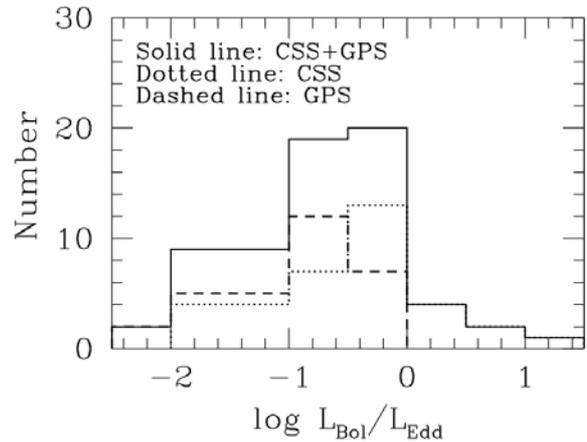

**Figure 2** The histogram distribution of the Eddington ratios.

### 2.2 Accretion/jet activity and O III kinematics

Figure 3 plots BH mass, $M_{bh}$, versus velocity dispersion, $\sigma_{\text{[O III]}}$ (=FWHM(O III)/2.35), and includes 23 CSS sources (squares), 5 GPS sources (triangles) and 9 linear radio sources (circles). The solid line is the relation of Tremaine et al. (2002)[33] for 31 nearby inactive galaxies. It is clear that both CSS sources and linear radio sources systematically deviate from the Tremaine et al. relation. Two of the five GPS sources also deviate substantially (OQ 172 and 4C 12.50).

We calculate the deviation $\Delta\sigma = \sigma_{\text{[O III]}} - \sigma_{\text{[pred]}}$ for our sample, where $\sigma_{\text{[pred]}}$ is calculated from the Tremaine et al. relation with the estimated BH mass. Wu (2009a)[22] have shown that there is no relation between the deviation of [O III] emission line $\Delta\sigma$ and the jet power $Q_{jet}$ in young radio galaxies. Therefore, the deviation of the [O III] emission line may not be directly caused by the jet power, and differences in the interstel-lar medium and/or other physical reasons may also play an important role in shaping the kinematics of the narrow emission lines. Figure 4 plots $\Delta\sigma$ versus Eddington ratio $L_{bol}/L_{Edd}$ for our sample (20 CSS + 3 GPS sources + 9 linear radio sources). For comparison with the radio quiet AGN (where the jet is weak, if present), we also calculate $\Delta\sigma$ and the Eddington ratio $L_{bol}/L_{Edd}$ for 31 NLS1s (circles) and 49 QSOs (crosses). The NLS1s are selected from Grupe & Mathur (2004)[17] and the QSOs are selected from Shields et al. (2003)[34], and BH masses are calculated from $L_{5100}$ and the broad line width, and bolometric luminosities are calculated from $L_{5100}$[e.g.,35]. We find that the distribution, $\Delta\sigma$-$L_{bol}/L_{Edd}$, in young radio galaxies (GPS/CSS) and linear radio sources is similar to that of radio quiet AGN, where the sources with $L_{bol}/L_{Edd}$~1 have the largest deviations (see Figure 4).

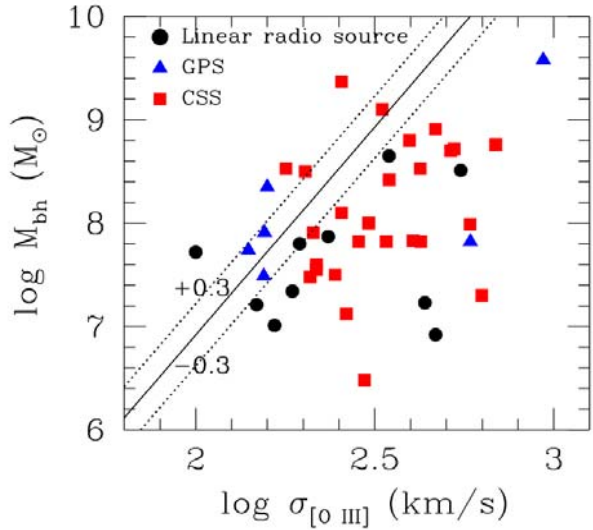

**Figure 3** The velocity dispersion $\sigma_{[O\,III]}$ vs. BH mass $M_{bh}$. The solid line is the $M_{bh}$-$\sigma$ relation of Tremaine et al. 2002 for 31 nearby inactive galaxies, and the dotted lines are their intrinsic scatter 0.3 dex.

The Eddington ratio, which is proportional to the accretion rate per unit BH mass, characterizes the extent to which radiation pressure competes with gravity in the nucleus, and may also play an important role in shaping the kinematics of the [O III] emission line, in addition to the influence of the gravitational potential of the bulge [e.g.,19]. After removing the blue wings, the core component of $\sigma_{[O\,III]}$ is still a good proxy of stellar velocity dispersion $\sigma$ [e.g.,19]. Blue wings are always found in NLS1s with high Eddington ratios[e.g.,20], which may result from the winds/outflows driven by the strong radiation pressure in these high Eddington ratio sources. It is interesting to note that both the young radio sources and the linear radio Seyfert galaxies with Eddington ratios $L_{bol}/L_{Edd}$~1 also have the largest deviations, which are similar to those of radio quiet NLS1 galaxies (see Figure 4). Therefore, accretion activity may still play an important role in shaping the kinematics of [O III] narrow emission line in the young radio galaxies, in addition to the possible effects of jet-ISM interactions.

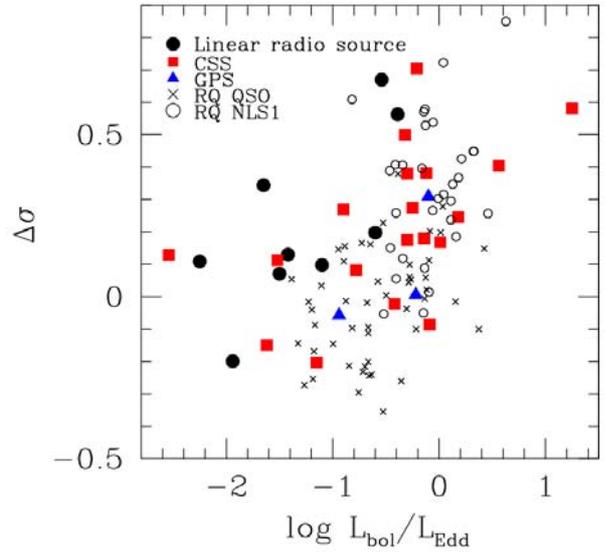

**Figure 4** The relation between the Eddington ratio $L_{bol}/L_{Edd}$ and $\Delta\sigma$.

## 3. Summary

Using a sample of 81 young radio galaxies (42 CSS + 39 GPS sources) selected from the literature, we estimate the BH masses, Eddington ratios, and investigate the possible physical reasons for deviations of width of [O III] narrow emission line. The main results can be summarized as follows:

(1) The BH masses in young radio galaxies range from $10^7$-$10^{10}M_{sun}$, and the mean value is <log $M_{bh}$>~8.3 (see Figure 1), which is less than that of radio loud QSOs or low redshift radio galaxies (<log $M_{bh}$>~9.0).

(2) Most of the CSS/GPS sources have relatively high Eddington ratios, with mean value <log $L_{bol}/L_{Edd}$> = -0.75 (see Figure 2), which are similar to those of NLS1s. This result supports the idea that the young radio galaxies are not only in the early stage of jet activities but also in an early stage of accretion activities.

(3) The $M_{bh}$-$\sigma_{[O\ III]}$ relation in young radio galaxies systematically deviates that defined by nearby inactive galaxies, and reminiscent of NLS1s (see Figure 3). The similarities between the young radio galaxies and radio quiet QSOs/NLS1s in $\Delta\sigma$-$L_{bol}/L_{Edd}$ distribution suggest that accretion activities may still play an important role in shaping the [O III] kinematics in these young radio galaxies (and also linear radio sources), since that radio jet is weak or absent in radio quiet AGN.

Astrophys J, 2003, 629: 61-71